\documentclass[doublecol]{epl2}
\usepackage{scalefnt}

\title{Nonextensivity in the Solar Neighborhood}
\shorttitle{Nonextensivity in the Solar Neighborhood}

\author{D. B. de Freitas\inst{1,2}\thanks{E-mail:
\email{danielbrito@dfte.ufrn.br}}
\and J. R. De Medeiros\inst{2}\thanks{E-mail: \email{renan@dfte.ufrn.br}} }
\shortauthor{D. B. de Freitas and J. R. De Medeiros}

\institute{
	\inst{1}Instituto de Educa\c{c}\~ao, Ci\^encia e Tecnologia do Rio Grande
do Norte, 59550-000
    Jo\~ao C\^amara, RN, Brazil\\
  \inst{2} Departamento de F\'{\i}sica,
    Universidade Federal do Rio
    Grande do Norte, 59072-970
    Natal,  RN, Brazil\\
}

\abstract{
In the present study, we analyze the radial velocity distribution as a
function
of different stellar parameters such as stellar age, mass, rotational
velocity and
distance to the Sun for a sample of 6781 single low--mass field dwarf stars,
located in the solar neighborhood. We show that the radial velocity
distributions are best fitted by $q$--Gaussians that arise within
the Tsallis nonextensive statistics. The obtained distributions cannot be
described by the standard Gaussian that emerges within
Boltzmann-Gibbs (B--G) statistical mechanics. The results point to the
existence of a hierarchical structure in phase space, in contrast to the
uniformly occupied phase space of B--G statistical mechanics, driven by the
$q$--Central Limit Theorem, consistent with nonextensive statistical
mechanics.}

\begin{document}

\maketitle

\section{Introduction}

The behavior of the distribution of different stellar physical parameters,
in particular rotational and radial velocities, seems to be better explained
on the basis of a Tsallis maximum entropy distribution function
\cite{tsallis1988} than by
standard Gaussian and Maxwell--Boltzmann distributions or analytical
functions (Soares {\it et al.} \cite{Soares06} and Carvalho {\it et al.}
\cite{carvalho2007, carvalho2008, carvalho2009}). Indeed, the nonextensive
generalization of Boltzmann-Gibbs statistics, presented by Tsallis
\cite{tsallis1988},
appears now as a powerful parametrization of the statistical mechanics
of out-of-equilibrium states in open systems as is the case of stars,
planetary
and stellar systems.

The Tsallis statistics, is based on $q$--exponential and
$q$--logarithm functions defined by,

\begin{equation}
\label{tsallisa}
\exp_{q}(f)=[1+(1-q)f]^{1/1-q},
\end{equation}
\begin{figure}
\centering
\resizebox{0.48\textwidth}{!}{%
  \includegraphics{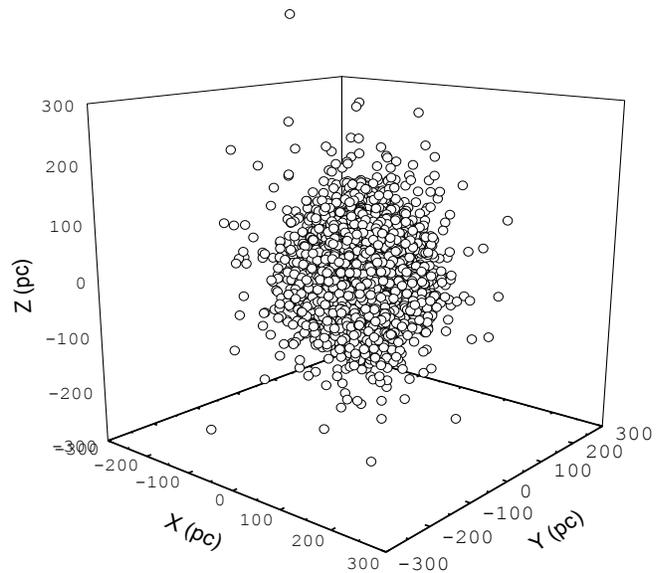}
  }
\caption{The spatial distribuition of the present working sample of
dwarf stars in the solar neighbourhood. The Sun is centered in point
(0,0,0).}
\label{fig0}       
\end{figure}
\begin{equation}
\label{tsallisb}
\ln_{q}(f)=\frac{f^{1-q}-1}{1-q},
\end{equation}
from where, associated with the $q$--statistics emerges the entropy
$S_{q}$ \cite{gell2004}

\begin{equation}
\label{tsallis1}
S_{q}=\frac{1-\sum_{1}p^{q}_{i}}{q-1} \quad\ (q\in \texttt{R}),
\end{equation}
where the $q$--Gaussian recovers the usual one at $q=1$. This new
distribution
function, the $q$--Gaussian,
was applied successfully to different astrophysical problems
\cite{hamity1996,lavagno1998,taruya2005,boghosian1996,zanette1995,kania1996,rajap1996,taruya2002,lima2002,defreitas09},
in addition of stellar radial and rotational velocities.

\begin{figure*}
\centering
\resizebox{0.68\textwidth}{!}{%
  \includegraphics{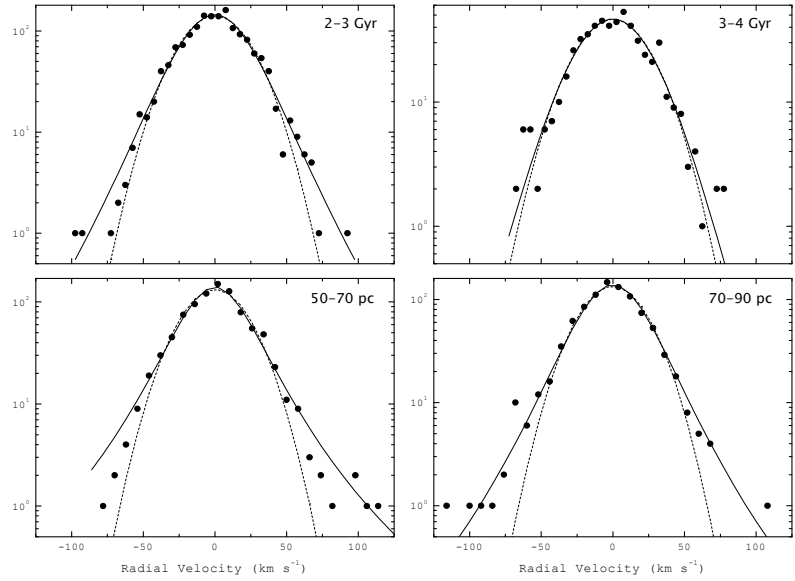}
  }
\caption{Comparison between the semi-log plot of the observed (dots) and fitted Gaussian (dashed line) and $q$-Gaussian (full line) distribution of radial velocity of F-type stars. Upper and lower panels represent, respectively, examples of the distributions segregated by age and distance to the Sun.}
\label{figa}       
\end{figure*}

In the present work we study the behavior of the distribution of stellar
radial
velocity in the solar neighborhood, using the Tsallis maximum entropy
distribution. This work brings the analysis, for
the first time, to our knowledge,
of the behavior of the $q$--Gaussian distribution applied to a stellar
parameter as a function of other additional relevant stellar parameters.
In fact, we study the behavior of the
$q$--Gaussian distribution of
radial velocity as a function of stellar age, mass, rotation and distance
of the stars to the Sun.
This new investigation also offers the possibility of check the validity
of the $q$--Central Limit Theorem,
recently conjectured by Umarov, Tsallis and Steinberg \cite{umarov2008}
for physical sub-systems, such as those associated to stars and stellar
systems.

In section 2 we describe the sample and how we selected stars and physical
stellar parameters for the present analyses. Section 3 brings the main
results of the present analysis. Finally, conclusions are presented in
last section.

\begin{figure*}
\centering
\resizebox{0.68\textwidth}{!}{%
  \includegraphics{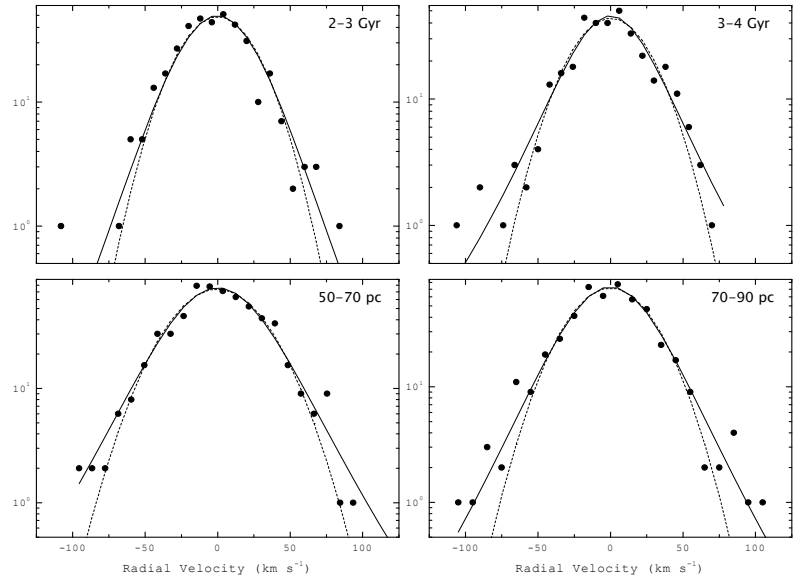}
  }
\caption{Comparison between the semi-log plot of the observed (dots) and fitted Gaussian (dashed line) and $q$-Gaussian (full line) distribution of radial velocity of G-type stars. Upper and lower panels represent, respectively, examples of the distributions segregated by age and distance to the Sun.}
\label{figb}       
\end{figure*}

\section{Stellar Working Sample and Observational Data}\label{data}

The present working sample is composed of 6781 single F and G spectral types
dwarf stars,
with precise radial velocity computed by Nordstrom {\it et al.}
\cite{nordstrom04}.
According these authors the sample is complete to the magnitude limit of 7.8
and should also be volume complete for the F-- and G--type dwarfs stars to
a distance
of $\sim$40 pc. The referred sample is defined in a volume around the Sun
of $\sim$200 pc, as shown in fig. \ref{fig0}.

For our analysis we taken the radial velocities, mass, age and projected
rotational velocities
from Holmberg {\it et al.}\cite{holmberg2007}, which gives an improved
version
of the work by Nordstrom {\it et al.} \cite{nordstrom04}, with a new
calibration
for all the relevant stellar parameters. All the selected objects are
low-mass stars with effective
temperature ranging from 4500 K to 7800 K ($T_{\odot}\approx$ 5778
K)\footnote{$\odot\equiv Sun$},
presenting a mass range of 0.65$M_{\odot}$ to 2.0$M_{\odot}$ with error
estimation of about
0.05$M_{\odot}$ and metallicity range [Fe/H] between --0.75 and 0.5.
Nordstrom {\it et al.} \cite{nordstrom04} show that the distribution of
their estimated metallicities obeys a Gaussian distribution, with a mean
of --0.14 and dispersion of 0.19 dex. For futher details on the
observational procedure, calibration, and error analysis, the reader is
referred to \cite{nordstrom04,holmberg2007}.

\section{Results and Discussion}\label{results}

In the following, we analyze the behavior of the distribution of the
radial velocity
in the context of the Tsallis statistics, taking into account, in
particular, the
dependence of such a behavior on different stellar parameters, namely age,
mass, projected rotational velocity and distance of the stars to the Sun.
The generalized $q$-Gaussian distribution function for radial velocity
$v_r$ can be
written as ({\it e.g.}, \cite{carvalho2007})

\begin{table}[tbp]
\scalefont{0.5}
\centering
\caption{Best {\it q} and {\it $\sigma_q$}--values determined using
non-linear regression L--M method for all the stars of the present working
sample.}
\label{tab1}
\renewcommand{\arraystretch}{1.5}
\begin{tabular}{lccccc}
\hline \hline
Age(Gyr) & $N_t$ & $\sigma_{Gauss}$ & $q$  & $\sigma_q$ & $R^{2}$  \\
\hline \hline
0--1 & 386 & 32.78 & 1.37$\pm$0.37 & 28.07  & 0.97        \\
1--2 & 1701  & 28.31  & 1.23$\pm$0.22 & 25.87  & 0.99        \\
2--3 & 1939  & 31.24  & 1.26$\pm$0.21 & 28.10  & 0.99        \\
3--4 & 902  & 34.59  & 1.1$\pm$0.18 & 33.39  & 0.99        \\
4--5  & 452 & 32.22 & 1.56$\pm$0.03 & 25.09 & 0.96        \\
5--6 & 402 & 32.04 & 0.75$\pm$0.24 & 44.98  & 0.97        \\
6--7   & 325 & 33.37 & 1.25$\pm$0.47 & 38.89  & 0.95        \\
7--8 & 291  & 33.74  & 1.46$\pm$0.0.39 & 32.76 & 0.97        \\
8--9 & 228  & 34.74  & 1.17$\pm$0.62 & 43.18 & 0.92        \\
9--10 & 155  & 36.45  & 1.65$\pm$0.75 & 35.95 & 0.91        \\
\hline
\end{tabular}
\end{table}

\begin{table}[tbp]
\scalefont{0.5}
\centering
\caption{Best {\it q} and {\it $\sigma_q$}--values determined using
non-linear regression L--M method for the F-Type stars of the present
working sample.}
\label{tab2}
\renewcommand{\arraystretch}{1.5}
\begin{tabular}{lccccc}
\hline \hline
Age(Gyr) & $N_t$ & $\sigma_{Gauss}$ & $q$  & $\sigma_q$ & $R^{2}$  \\
\hline \hline
0--1 & 200 & 29.82 & 1.43$\pm$0.45 & 24.72  & 0.92        \\
1--2 & 1573  & 27.96  & 1.35$\pm$0.27 & 24.19  & 0.97        \\
2--3 & 1571  & 30.78  & 1.26$\pm$0.21 & 27.72  & 0.98        \\
3--4 & 560  & 33.75  & 1.1$\pm$0.08 & 32.50  & 0.97        \\
4--5  & 217 & 33.01 & 1.57$\pm$0.54 & 25.48 & 0.88        \\
5--6 & 146 & 43.21 & 0.89$\pm$0.67 & 44.82  & 0.78        \\
>6   & 156 & 30.20 & 1.29$\pm$0.33 & 40.04  & 0.97        \\
\hline
\end{tabular}
\end{table}

\begin{table}[tbp]
\scalefont{0.5}
\centering
\caption{Best {\it q} and {\it $\sigma_q$}--values determined using
non-linear regression L--M method for the G-Type stars of the present
working sample.}
\label{tab3}
\renewcommand{\arraystretch}{1.5}
\begin{tabular}{lccccc}
\hline \hline
Age(Gyr) & $N_t$ & $\sigma_{Gauss}$ & $q$  & $\sigma_q$ & $R^{2}$  \\
\hline \hline
0--1 & 186 & 34.65 & 1.51$\pm$0.66 & 27.97  & 0.93        \\
1--2 & 128  & 30.38  & 0.93$\pm$0.13 & 31.0  & 0.87        \\
2--3 & 368  & 33.24  & 1.18$\pm$0.43 & 31.04  & 0.96        \\
3--4 & 342  & 34.31  & 1.37$\pm$0.03 & 29.46  & 0.94        \\
4--5  & 235 & 33.32 & 1.43$\pm$0.43 & 27.98 & 0.97        \\
5--6 & 256 & 38.91 & 0.84$\pm$0.71 & 41.23  & 0.92        \\
6--7   & 247 & 46.1 & 1.16$\pm$0.47 & 43.35  & 0.93        \\
7--8 & 240  & 31.34  & 1.49$\pm$0.19 & 30.17 & 0.96        \\
8--9 & 210  & 34.17  & 1.24$\pm$0.66 & 40.67 & 0.91        \\
9--10 & 146  & 36.97  & 1.78$\pm$0.83 & 34.12 & 0.89        \\
\hline
\end{tabular}
\end{table}
\begin{equation}
\label{fi}
\Phi_q (v_r)=A_q \left[1-(1-q){v_r^2\over
\sigma_q^2}\right]^{1\over
1-q},
\end{equation}
where $\sigma_{q}$ is the characteristic width and $q$ is a free parameter
that denotes the entropic index. The  parameter $q$ is related to the size
of the tail
in the distribution ({\it e.g.},
\cite{defreitas09}). At the limit, when $q\rightarrow1$, the above
distribution reproduces the usual Gaussian distribution, given by
\begin{equation}
\label{gau}
\phi(v_r)=A \exp{\left(-{v_r^2\over \sigma^2}\right)}.
\end{equation}

\begin{table*}[tbp]
\scalefont{0.5}
\centering
\caption{Best {\it q} and {\it $\sigma_q$}--values determined using
non-linear regression L--M method for the F- and G-Type dwarf stars for
other stellar parameters.}
\label{tab4}
\renewcommand{\arraystretch}{1.5}
\begin{tabular}{l | ccc | ccc}
\hline
Parameters &  & F--Type &   &  & G--Type &  \\
\hline \hline
Massa(M$_{\odot}$) & $q$ & $\sigma_{q}$ & $R^{2}$  & $q$ & $\sigma_q$ &
$R^{2}$ \\
\hline \hline
0.90--1.10 & 1.33$\pm$0.31 & 31.31 & 0.98 & 1.43$\pm$0.03  &  30.16 &
0.99    \\
1.10--1.30 & 1.21$\pm$0.08  & 27.52  & 1.0 & 1.48$\pm$0.61  & 31.32  &
0.91    \\
1.30--1.50 & 1.27$\pm$0.13 & 26.95 & 1.0 & 0.66$\pm$0.59  &  39.71 &  0.94
   \\
1.50--2.00 & 1.29$\pm$0.33  & 26.15  & 0.97 & 1.23$\pm$0.1  & 33.38  &
0.88    \\
\hline
average & 1.28$\pm$0.21  & --  & -- & 1.2$\pm$0.33  & --  &  --    \\
\hline
$v$sin$i$(km s$^{-1}$) & $q$ & $\sigma_{q}$ & $R^{2}$  & $q$ & $\sigma_q$
& $R^{2}$ \\
\hline\hline
1--5 & 1.2$\pm$0.21 & 34.55 & 0.99 & 1.29$\pm$0.13  &  34.95 &  0.99    \\
5--10 & 1.22$\pm$0.31  & 31.44  & 0.99 & 1.45$\pm$0.45  &  26.13 &  0.96
 \\
10--20 & 1.13$\pm$0.20 & 27.5 & 0.99 & 0.59$\pm$1.37  & 34.42  &  0.87    \\
20--40 & 1.28$\pm$0.23  &  24.74 & 0.99 & 1.62$\pm$0.30  &  24.71 &  0.87
  \\
$>$40 & 1.15$\pm$0.18 & 26.36 & 0.99 & --  &  -- &  --    \\
\hline
average & 1.20$\pm$0.23  & --  & -- & 1.24$\pm$0.6  & --  &  --    \\
\hline
R$_{*,\odot}$(pc) & $q$ & $\sigma_{q}$ & $R^{2}$  & $q$ & $\sigma_q$ &
$R^{2}$ \\
\hline \hline
0--30 & 1.29$\pm$0.67 & 24.44 & 0.89 & 2.0$\pm$0.32  &  20.72 &  0.99    \\
30--50 & 1.29$\pm$0.30  & 29.11  & 0.96 & 1.23$\pm$0.42  & 34.80  &  0.96
  \\
50--70 & 1.46$\pm$0.15 & 24.56 & 0.99 & 1.26$\pm$0.33  &  36.42 &  0.97    \\
70--90 & 1.34$\pm$0.17  & 25.92  & 0.99 & 1.26$\pm$0.01  & 33.50  &  0.97
  \\
90--110 & 1.01$\pm$0.01  & 33.91  & 1.00 & 1.42$\pm$0.67  & 29.36  &  0.91
   \\
110--130 & 1.40$\pm$0.28  & 24.65  & 0.98 & 1.30$\pm$0.71  & 30.59  &
0.93    \\
>130 & 1.33$\pm$0.27  & 25.73  & 0.98 & 0.80$\pm$2.21  & 36.70  &  0.65    \\
\hline
average & 1.30$\pm$0.26  & --  & -- & 1.32$\pm$0.67  & --  &  --    \\
\hline
\end{tabular}
\end{table*}

\begin{figure}
\centering
\resizebox{0.40\textwidth}{!}{%
  \includegraphics{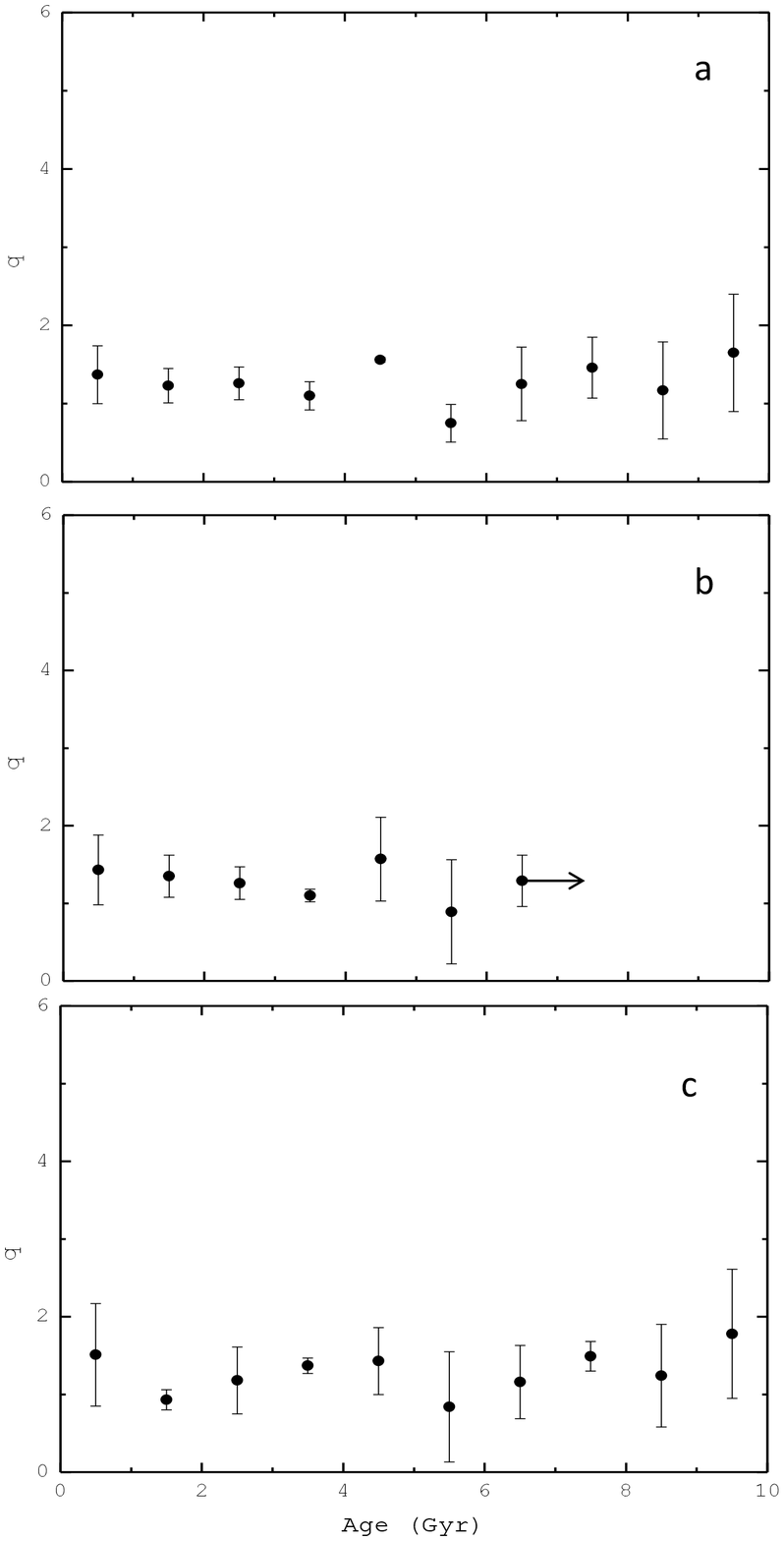}}
\caption{Values of the fitted parameter $q$ as a function of the stellar
age for a) all stars, b) only F-Type stars and c) only G-Type stars of the
present working sample.}
\label{fig3b}       
\end{figure}

All $q$-Gaussian distributions are two-parameters ($q - \sigma_{q}$)
nonlinear functions given by eq. (\ref{fi}). The result of our
computation
for these two parameters, for well defined stellar age intervals is presented
in tables \ref{tab1}, \ref{tab2} and \ref{tab3}, which brings also the
total number
$N_{t}$ of stars in each age interval, $\sigma_{Gauss}$ of the Gaussian
distribution from eq. (\ref{gau}) and $R^{2}$--parameter obtained by
a nonlinear regression method based on the Levenberg-Marquardt algorithm
\cite{leven,marqu}. This method was used to compute the $q$-Gaussian with
symmetric Tsallis distribution from eq. (\ref{fi}). The obtained values
of the parameter $q$, strongly suggest that the distribution of radial
velocity for
the present sample of field dwarf stars is far from being
in agreement with a standard Gaussian, since the values of this parameter
are significantly
different from 1, independent of the stellar age considered, as well as
the considered
spectral types. The error bars in fig. \ref{fig3b} correspond to a 0.05
confidence
limit, as given in tables \ref{tab1}, \ref{tab2} and \ref{tab3}.

To illustrate the results listed in the referred tables, we present in 
figs. \ref{figa} and \ref{figb} a few exemples of
the obtained fits for the distribution of stellar age and distance to 
the Sun, once we consider the usual and the $q$-Gaussian.
As we can see, the $q$-Gaussian is the more suitable distribution fitting 
the obtained functions, irrespective of the region
of the distributions, in contrast to the standard gaussian that fits 
very well only the central regions of the observed distributions.
Such a behavior is the same, irrespective of the stellar parameter here 
considered.

For instance, Carvalho {\it et al.} \cite{carvalho2007} have found that
for stellar clusters
older than 1 Gyr exists a positive gradient between the parameter $q$ of
radial velocities
distributions and the stellar age, while for young clusters ($<$1 Gyr)
this correlation is small. From fig. \ref{fig3b} one observes that the values of $q$ do not change
whether we considered the whole of the
stellar sample or the stars segregated by spectral types. In general, for
the three cases shown in fig. \ref{fig3b}, the values of the $q$ are
roughly constant over the stellar age and are consistent with the average
value 1.28. Such a behavior for $q$ is very similar to the results
obtained by
Carvalho {\it et al.} \cite{carvalho2007} for young stellar clusters.
Interestingly,
the great majority of the young clusters analyzed by those authors are
located in the
solar neighborhood near the galactic disc, namely at the same region where are
located the stars
of the present sample.

In addition, we have analyzed the behavior of the $q$-parameter, obtained
from
de distribution of the radial velocity, as a function of stellar mass,
projected rotational velocity and distance of the stars to the Sun.
Table \ref{tab4} summarizes the results from such an analysis, from where
one can observe
that the values of $q$ are systematically different of 1, irrespective of the
stellar parameter considered, in reinforcing that the Tsallis function fit
the
observed stellar
residual velocity better than the usual Gaussian does.

\section{Conclusions}\label{conclusions}

We used $q$-Gaussian distributions to investigate the radial velocity of a
sample of 6781 single field dwarf stars of spectral types F and G in the
solar neighborhood. These low-mass stars have ages ranging between 1 and
10 Gyr and distance to the Sun from a few to tens of parsecs. In addition to
the age and distance to the Sun, we studied the behavior of the
distribution of the radial
velocity as a function of two other relevant parameters, namely stellar
mass and
projected rotational velocity. The present analysis shows that radial
velocity
distribution for these stars is far from being a standard Gaussian model.
The $q$-Gaussian distribution described in Tsallis nonextensive statistics
is clearly the best distribution fitting the obtained functions.

The relationship observed between the $q$-parameter and stellar
age, stellar mass, rotational velocity and distance to the
Sun indicates that the significant radial velocity deviation of the
stellar sample from a standard Gaussian is a result of statistical
dependence between the distributions associated with these variables. On
the other hand, this dependence reveals the effects of long-range interactions and the
formation of high-energy tails consistent with the $q$-CLT, where the
nonextensive character is observed.

\acknowledgments
Research activity of the Stellar Board of the Federal University of Rio
Grande do Norte (UFRN) and at the Federal Institute of Rio Grande do Norte
(IFRN)
are supported by continuous grants from FNDE (PET-Programa de
Educa\c{c}\~ao Tutorial) and FAPERN brazilian agency. We also acknowledge
continuous financial support from CNPq brazilian.

\end{document}